# Design and validation of an index to measure development in rural areas through stakeholder participation

Abreu, I.[a], Mesías, F.J.[a]*, Ramajo, J.[a]

[a] Department of Economics- University of Extremadura. Avda. Adolfo Suarez, s/n – 06007 Badajoz (Spain)

*Corresponding author: Francisco J. Mesias; email: fjmesias@unex.es; Tel.: (0034) 924289300; Fax: (0034)924286201. ORCID iD: 0000-0001-5334-9554

## ABSTRACT

In the context of the discussion of the European Union's Multiannual Financial Framework 2021-2027, and with Rural Development (RD) being the Second Pillar of the Common Agricultural Policy, it's urgent to evaluate Rural Development policies despite their lower weight compared to agricultural and agri-food activities. Its multi-dimensional characteristic - dealing with such diverse aspects as employment, modernization, sustainability, or environment - makes it essential to use a composite index made up of different indicators, whose number doesn't compromise the calculation and interpretability of the index.

This paper therefore proposes the development of an index to assess rural development based on a set of 25 demographic, economic, environmental, and social welfare indicators previously selected through a Delphi approach. Three widely accepted aggregation methods were then tested: a mixed arithmetic/geometric mean without weightings for each indicator; a weighted arithmetic mean using the weights previously generated by the Delphi panel and an aggregation through Principal Component Analysis.

These three methodologies were later applied to 9 Portuguese NUTS III regions, and the results were presented to a group of experts in rural development who indicated which of the three forms of aggregation best measured the levels of rural development of the different territories. Finally, it was concluded that the unweighted arithmetic/geometric mean was the most accurate methodology for aggregating indicators to create a Rural Development Index

## 1. INTRODUCTION

In a world that is increasingly developed and urbanized and where population movements from the rural areas to the cities have been taking place for decades, the rural environment has found itself in an increasingly unfavourable situation with respect to the urban areas (Abreu et al., 2019). Not only this "emptying" of villages deprives them of part of their most active and dynamic population, but also the lack of population itself leads to the disappearance of essential services, which in turn encourages rural exodus.

This is reflected in an increasingly pronounced imbalance between the level of development of rural and non-rural areas. Thus, indicators such as the risk of poverty or social exclusion (23.9% in rural areas vs. 21% in urban areas), the level of education (60-80% of city-dwellers have tertiary education, but less than 40% in most rural areas) or the level of digital skills among adults (49% adults have basic or higher skills in rural areas vs. 63% of those living in cities) show worse values in rural areas than in urban areas (Eurostat, 2019), aspects that clearly influence the job opportunities and life development of citizens.

Moreover, there is a growing disconnection between urban and rural dwellers, with the former ceasing to see the latter as a necessary and fundamental part of society. This leads to a situation where "urban" citizens perceive that they are subsidizing the "rural" ones, without getting any benefit in return. Thus, they ignore, out of unawareness derived from urban life, the valuable environmental, cultural, and social services that the villages and their inhabitants have provided and continue to provide to their fellow citizens in the cities.

It is within this framework that the concept of Rural Development (RD) arises, which could be defined as the set of initiatives aimed at fostering the modernization of rural areas, the creation of new job opportunities, the sustainability and efficiency of farms and the preservation of ecosystems (Abreu and Mesias, 2020; UPA, 2016). Rural development has been promoted for decades in different parts of the world, such as the European Union, where the Rural Development Policy, also known as the Second Pillar of the Common Agricultural Policy, has been gaining importance over agricultural policies, even though agriculture and agri-food activities are still a major component of RD policies (European Commission, 2017). Like any other policy, RD needs to be evaluated to determine the effectiveness of the measures implemented, to analyse new initiatives or to decide which areas need more attention (Abreu et al., 2019). This process, which is complex for any policy, is even more so in the case of rural development, because, as it has been mentioned, it deals with such diverse aspects like employment, modernization, sustainability, and environment.

Several indices have been created to try to measure both development in general terms and rural development specifically. Among the former we can highlight the Gross Domestic Product (World Bank, 1997), the Human Development Index (HDI) (UNDP, 2016), the Social Development Framework (Davis, 2004) or the Multidimensional Poverty Index (Oxford Poverty and Human Development Initiative, 2010). However, none of them is specifically designed for the assessment of rural areas. Therefore, several authors have tried to develop specific indices to measure rural development, such as the one by Kageyama (Kageyama, 2008), which was applied, for example, to assess the effectiveness of public policies in Brazil (Haag, 2009) or the one by Abreu (Abreu, 2014) which measured RD in different Portuguese municipalities (Abreu et al., 2019).

Since RD is a multi-dimensional process, RD indexes should be based on the use of several indicators representing its different dimensions. Moreover, given that the implementation of a policy does not only affect the targeted activity but also contributes to the modification of the entire human environment, it cannot be monitored merely by verifying compliance with its objectives (Carraro et al., 2009).

An indicator is a quantitative or a qualitative measure derived from a series of observed facts that can reveal relative positions (e.g., of a country) in a specific area. Thus, when evaluated at regular intervals it can point out the direction of change across different units and through time. A composite index is formed when individual indicators are compiled into a single index on the basis of an underlying model (OECD, 2008).

For this reason, policy analysts and policy makers turn to composite indexes, which are better equipped to capture the different and multi-dimensional natures of development. If composite indexes are to measure progress, they need to incorporate enough indicators so that multidimensionality is captured without compromising the interpretability of the index. Hence, the selection of underlying indicators is the result of a trade-off between possible redundancies caused by overlapping information and the risk of losing information (Kynčlová et al., 2020).

Ideally, the composite RD index should measure multi-dimensional concepts that cannot be captured by partial indicators alone and should therefore embrace all the most important rural development domains (DEFRA, 2004). While the main areas of policy concerns related to rural development have been relatively easily identified, i.e.: i) economic structure and performance, ii) social well-being and equity, iii) population and demographics, and iv) environment and sustainability, overcoming these constraints in individual rural areas through precise targeting of policy interventions has proven to be a complex policy task, mostly due to their local/regional specificity as well as complex links among individual growth components and their constraints (Michalek and Zarnekow, 2012a).

In this context, it can therefore be noted that the selection of the indicators included in the RD index is as relevant, if not more, than the way in which they are combined. Thus, although indices such as those of Kageyama (2008), Abreu (2014) or Michalek and Zarnekow (2012b) use different indicators related to economic, demographic, social and environmental aspects, all of which are highly relevant for RD, they all suffer from a selection of indicators based on the literature and not on what stakeholders in rural areas consider to be really relevant for their development. Furthermore, the calculation of the values obtained by these indices is based on different methodological approaches whose validity has not been tested.

Therefore, and to overcome these limitations, this paper proposes the development of a RD index using the indicators selected by a panel of experts through a Delphi approach (Abreu and Mesias, 2020). Subsequently, different aggregation methods accepted by the scientific community are used to build the RD index: simple aggregation of indicators through geometric/arithmetic mean, as used on the HDI (Conceição, 2019) or on the RD Index proposed by (Abreu et al., 2019); weighted arithmetic average of the indicators using the results of a panel of RD experts through a Delphi approach (Abreu and Mesias, 2020); Principal Component Analysis (PCA), widely used by different authors such as (Bolcárová and Kološta, 2015; Jollife and Cadima, 2016; Yilmaz et al., 2010). The 3 methodologies have been then applied to 9 different Portuguese NUTS III territories (a lower level than the regional, but at which RD policies are applied by countries), using the most recent data available for the whole set of variables from the National Statistical Institute of Portugal.

At a final stage, the results of the different RD indexes were revised by a panel of Portuguese experts on RD, who identified the most accurate methodology for aggregating the indicators when creating a Rural Development Index (RDI). This research therefore provides an effective tool for both policymakers and managers of RD programs, which due to its composite nature can be applied to analyse the main determinants of rural/regional development in individual rural areas. Furthermore, it can also be used to measure the impact of cohesion policy and RD/structural programmes at various regional levels.

## 2. MATERIALS AND METHODOLOGY

### 2.1. Data collection

As previously stated, the aim of this paper is to test different RD indexes and select the one which would provide the more comprehensive results. Therefore, the intention was not to obtain an evaluation of the level of rural development in an entire country or a comparison between different regions of that country, but to apply the RD indexes to various regions with substantial differences in levels of development so that the results provided could be compared.

In this context, and among the different administrative levels (NUTS I, NUTS II, NUTS III and LAU), it was chosen to apply the study at the NUTS III level in Portugal. The choice of the NUTS III level was due to this being the minimum level at which rural development policies are generally applied. In fact, the only exception is the Community-Led Local Development (CLLD), applied at a Local Administrative Unit (LAU) level by the Local Action Groups (LAG).

In administrative terms, in Portugal there are 3 NUTS I, 7 NUTS II and 25 NUTS III regions. For this paper, and in order to reflect the different realities of the country, it was decided to select a total of 9 NUTS III regions: 2 from each of the NUTS II regions Norte, Centro and Alentejo, plus the Algarve, Autonomous Region of Madeira and Autonomous Region of Açores (where the NUTS II and NUTS III levels coincide). Figure 1 presents the different NUTS III regions which have been used in this research.

**Figure 1. The nine NUTS III selected (Portuguese islands and continental Portugal)**

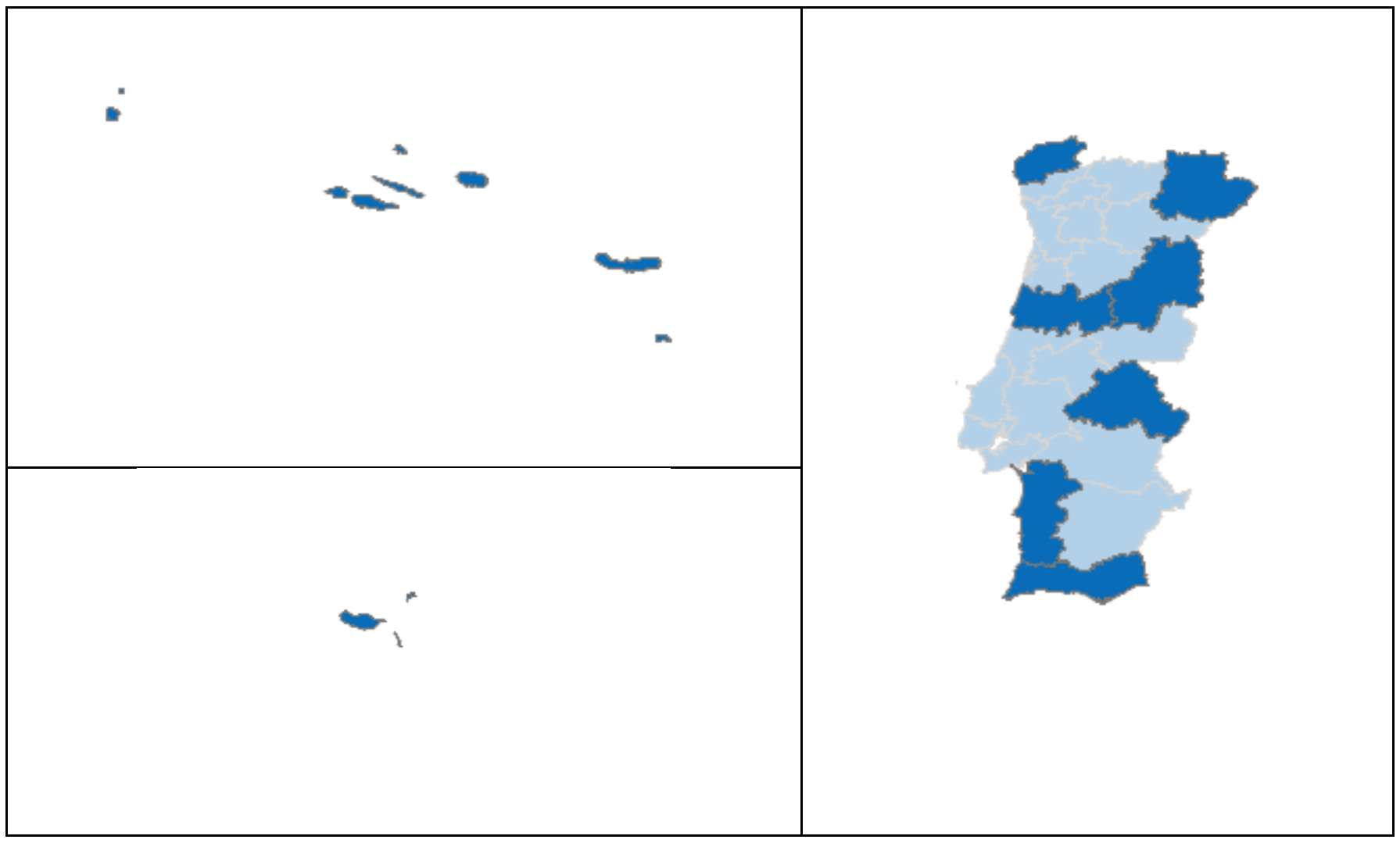

SOURCE: Own elaboration.

The indicators selected for this research were identified in a preliminary study by the authors (Abreu and Mesias, 2020). The Delphi qualitative methodology was used to objectively identify the indicators that should be included in a Rural Development Index that would allow the accurate monitoring and evaluation of RD policies and programs. Based on the results of a panel of experts with different roles in RD, 25 indicators were selected -from an initial group of 88- to describe the four basic dimensions/pillars of RD: population, social welfare, economy, and environment. The experts also assigned different weights to the selected indicators. Table 1 lists the selected indicators, together with the year of the data collected for this research.

**Table 1. Description of the indicators to be included in the index**

| | | Abbrev. | Data year |
|---|---|---|---|
| Population | Demographic Dependency Index (%) - ratio between those 65 and older plus those under 15 and the population in the working ages (ages 15-64) | DmgDep | 2019 |
| | Proportion of population aged 65 or over (%) | Pop65 | 2019 |
| | Proportion of population aged 16 or under (%) | Pop16 | 2019 |
| | Population density (inhabit/km$^2$) | PopDens | 2019 |
| | Rate of natural increase (%) | NatInc | 2019 |
| Socia | Coverage of essential health services (%) | HlthServ | 2019 |
| | Share of workforce with al least post-secondary education completed (%) | WorkQual | 2011 |
| | Literacy (%) – Proportion of the population aged 10 or more who can read or write | Lit | 2011 |
| | Proportion of youth and adults with ICT skills (%) | ICT | 2020 |
| | Share of university students (%) - Proportion of population in universities | Univ | 2011 |
| | Proportion of conventional dwellings of regular residence with facilities (%) | Facil | 2011 |
| | Proportion of population covered by a mobile network (%) | MobNet | 2011 |
| Economy | Average earnings per capita (€/inhabit) | Earn | 2018 |
| | Gross family income (€/year) | FamInc | 2018 |
| | Per capita purchasing power (%) | PurcPw | 2017 |
| | Unemployment rate (%) | Unemp | 2011 |
| | Total income primary[1] sector (Million €) | IncPrim | 2018 |
| | Total Gross Value Added of the primary sector (% of GDP) | PrimGVA | 2018 |
| | Research and development expenditure as a proportion of GDP (%) | R&D | 2017 |
| Environment | Renewable energy share in the total final energy consumption (%) | RenEn | 2011 |
| | Proportion of treated wastewater (%) | WasteW | 2009 |
| | Proportion of important sites for terrestrial and freshwater biodiversity that are covered by protected areas (%) | ProtectA | 2019 |
| | Proportion of bodies of water with good ambient water quality (%) | WatQlt | 2018[2] |
| | Proportion of agricultural area under productive and sustainable agriculture (%) | SustAgr | 2009 |
| | Total expenditure per capita spent on the preservation, protection and conservation of all cultural and natural heritage (€/inhabit) | ExpHer | 2019 |

SOURCE: Own elaboration, from Instituto Nacional de Estatística de Portugal and (Abreu and Mesias, 2020).

[1] Agriculture, Forestry, and Fisheries, aquaculture & fish processing.

2 Açores e Madeira: 2015. Continental Portugal: 2016-2018.

Data were directly collected from the official website of the National Statistical Institute of Portugal (INE) (www.ine.pt) and the Portuguese Communications National Authority (www.anacom.pt) and are the most recent data available. For some of the indicators retrieved from INE, and due to the type of data provided, some methodological adjustments had to be made as detailed below.

- *Coverage of essential health services*. As in Portugal there is a National Healthcare Service covering all the population, it is the quality of the service which can differ from one place to another. Therefore, a composite indicator resulting from the arithmetic average between the normalized values per 1,000 inhabitants of Medical doctors (No.) and Beds (No.) in hospitals was used.
- *Proportion of youth and adults with ICT skills*. As there are no Portuguese data available at NUTS III level, this value was estimated through the proportion of people aged between 16 and 74 years old using Internet in the 12 months prior to the interview by INE.

- *Proportion of bodies of water with good ambient water quality*. As there were no data available at NUTS III level for Açores and Madeira, an indirect source was used to obtain data for Açores (Governo Regional dos Açores, 2021). The figures for Madeira were then extrapolated from those of Açores since both NUTS III regions are very similar due to their island nature.

### 2.2. Aggregation methods

The variables presented above were subsequently combined to create RD Indexes through the application of different aggregation methods selected after a literature research and considering (or not) weightings for the indicators. Thus, in a first classification it is possible to find indexes in which simple aggregation of indicators is used through arithmetic or geometric averaging (Abreu, 2014; Kageyama, 2008), while other studies use PCA as an aggregation technique (Bolcárová and Kološta, 2015; Jollife and Cadima, 2016; Michalek and Zarnekow, 2012b). Another classification approach comes from the way of weighting the indicators: i) they can be assigned the same weight, as in the Human Development Index (HDI) (UNDP, 2010) or in several Rural Development Indexes (Abreu, 2014; Kynčlová et al., 2020); ii) the weighting itself is derived from the loading of the PCA factors; iii) a weighting specifically developed for this task -e.g. through a preliminary Delphi study, as in the case of (Abreu and Mesias, 2020)- can be used.

Within the first group, the Kageyama and Abreu indexes are quite similar since the former uses an arithmetic mean of all the indicators while the latter performs an arithmetic mean of the indicators within each pillar of the RD, and then aggregates the scores of each pillar through a geometric mean, thus avoiding the substitution effect of Kageyama’s arithmetic approach. For this reason, within this first type of indicators that use simple aggregation of indicators, it was decided to use Abreu's RD index (*RDI Abreu*) for this research, as it is considered an evolution of Kageyama's index. Additionally, two others indexes will also be analysed, one that uses the PCA for aggregation and weighting of indicators (*RDI PCA*) and finally another based on a weighted arithmetic average of the indicators, using the weights generated by the Delphi panel of experts and retrieved from (Abreu and Mesias, 2020) (*RDI Delphi*). The aggregation methodology in the three indexes is detailed below.

### *2.2.1. RDI Abreu*

*RDI Abreu* was proposed in 2014 (Abreu, 2014) and considers four different dimensions or sub-indexes: demography (Population Index), economy (Economic Index), social welfare (Social Welfare Index), and environment (Environment Index). Normalized values of the indicators are used and first it is calculated the arithmetic mean of the indicators belonging to each dimension. After that, the scores of the 4 dimensions are aggregated by means of a geometric mean, avoiding a substitution effect that would come from the use of arithmetic average. This is also the method used since 2010[1] in HDI. With the use of the geometric mean, a territory with significantly lower values in one dimension will have its RDI significantly penalized, instead of having its result biased by extreme values (as in arithmetic mean). The underlying concept is that one territory cannot be considered to have a high level of development if it has a poor performance in one of the dimensions of development. Therefore, the four dimensions (Population, Social, Economy and Environment) will have the same importance in the evaluation of a territory's Rural Development. As a drawback, this method considers that all the chosen indicators have the same weight/contribution to the final Index, which is not necessarily true.

### *2.2.2. RDI PCA*

When a composite index is to be constructed, PCA can be used when each pillar of the index is intended to describe a specific aspect of the latent phenomenon to be measured - in this study, Rural Development. In this case, the dimensions would be again Population, Social Welfare,

[1] Until 2010, HDI used the arithmetic mean to aggregate the dimensions (UNDP, 2016).

Economy and Environment, and within each of them the indicators would be allocated. These are considered as proxies and are related to that dimension and to each other, which justifies the use of the technique (Annoni and Dijkstra, 2019).

PCA is a statistical multivariate technique that transforms an original set of variables, initially correlated with each other, into a substantially smaller set of uncorrelated factors that contains most of the information in the original set. The underlying idea is simple: to reduce the dimensionality of a dataset, while preserving as much "variability" (i.e. statistical information) as possible or, in other words, finding new variables (the Principal Components) uncorrelated with each other and that are linear functions of the original variables (Jollife and Cadima, 2016).

A two-step procedure was followed in this piece of research: First, PCA was used to aggregate the indicators within each pillar; subsequently, a PCA was performed again using as inputs the ratings previously generated for each RD pillar, which are therefore weighted according to their own factor loadings (Bolcárová and Kološta, 2015).

In the first step, a maximum of 3 factors were considered for each dimension, depending on the number of factors needed to explain at least 80% of the variance of the dimension. The final percentages of variance explained for each dimension were: 96% for Population with 2 factors; 81% with 2 factors for Social Welfare; 88% with 3 factors for Economy; finally, 84% with 3 factors for Environment. The individual PCA pillar sub-indexes were then calculated by weighting the respective factors according to the variance explained by each one of them in its pillar.

To aggregate the values of the 4 sub-indexes obtained in the first stage into a single RDI for the 9 NUTS III studied, the PCA technique was used again with a maximum of two factors. The process concluded with an explanation of variability of 94%, 80% collected by the first factor and 14% by the second one.

Similar to Abreu's approach, the PCA method has also some weaknesses such as the fact that correlations do not necessarily represent the real influence of individual indicators on the phenomena being measured or the minimisation of the contribution of individual indicators which do not move with other individual indicators (OECD, 2008). Also the derivation of weights through PCA can be seen as neither straightforward nor transparent, because pure statistical approaches may lead to inappropriate normative results such as, for example, the assignment of negative weights to some dimensions (Decancq and Lugo, 2008).

#### *2.2.3. RDI Delphi*

(Abreu and Mesias, 2020) carried out a Delphi study in order to select the most relevant variables to be included in an RDI, as well as to obtain the weighting that each indicator should have in the final Index.

Out of 30 experts who took part in the study, more than 70% stated that the 4 dimensions proposed (Population, Social welfare, Economy and Environment) and their corresponding indicators should have different weights, thus differentiating their contribution to RD. Detailed values for each indicator and dimension can be found in (Abreu and Mesias, 2020). Nonetheless, and to summarize, we can indicate that the Economy pillar was considered the most relevant (28.4% weighting) followed by Social Welfare (26.2%), Environment (24%) and Population (21%).

In this study, *RDI Delphi* was calculated as a weighted arithmetic mean using the weights previously generated for each of the indicators by the Delphi experts panel (Abreu and Mesias, 2020).

### 2.3. Assessment by experts

The results generated by the three RDIs were then assessed by a panel of 25 Portuguese experts on Rural Development who had also previously taken part in the preliminary selection of the indicators (Abreu and Mesias, 2020). Experts had different backgrounds, although all of them were linked to RD and RD policies (13 worked for Local Action Groups, 5 in Public Administration, 5 were entrepreneurs and 2 were researchers). This number of experts is in line with other Delphi studies (Bélanger et al., 2012; Benitez-Capistros et al., 2014; Escribano et al., 2018; Horrillo et al., 2016). Experts were contacted by email and a summary of the results of the three RDIs for the nine NUTS III Portuguese regions analysed was sent to them. They were requested to indicate the index that, from their point of view, most accurately reflected the level of rural development of the different regions and to give some reasons supporting their answer.

## 3. RESULTS AND DISCUSSION

Table 2 presents the original values for each indicator prior to their aggregation in the RD indexes.

**Table 2. Values of each indicator for the NUTS III regions analysed**

| | | Alto Minho | Terras de Trás-os-Montes | Região de Coimbra | Beiras e Serra da Estrela | Alentejo Litoral | Alto Alentejo | Algarve | Região Autónoma dos Açores | Região Autónoma da Madeira |
|---|---|---|---|---|---|---|---|---|---|---|
| POPULATION | DmgDep | 57.60 | 67.10 | 60.00 | 64.40 | 62.80 | 63.60 | 58.40 | 43.50 | 43.00 |
| | Pop65 | 25.33% | 30.27% | 25.69% | 29.17% | 26.49% | 27.46% | 21.91% | 14.94% | 16.98% |
| | Pop16 | 11.22% | 9.90% | 11.81% | 10.01% | 12.09% | 11.41% | 14.95% | 15.37% | 13.11% |
| | PopDens | 103.80 | 19.40 | 100.10 | 33.60 | 17.60 | 17.20 | 87.70 | 104.60 | 317.20 |
| | NatInc | -0.65% | -1.00% | -0.56% | -1.09% | -0.59% | -1.15% | -0.16% | -0.06% | -0.31% |
| SOCIAL WELFARE | HlthServ | 285.14 | 368.97 | 927.47 | 347.18 | 164.13 | 315.20 | 323.67 | 325.79 | 400.58 |
| | WorkQual | 21.14% | 26.72% | 30.42% | 23.86% | 18.41% | 21.38% | 23.71% | 19.61% | 23.32% |
| | Lit | 96.76% | 95.93% | 97.12% | 95.68% | 94.52% | 94.45% | 96.40% | 96.43% | 95.00% |
| | ICT | 75.90% | 75.90% | 76.20% | 76.20% | 77.10% | 77.10% | 82.50% | 79.70% | 82.00% |
| | Univ | 11.36% | 13.82% | 17.81% | 12.58% | 10.28% | 11.21% | 13.51% | 10.77% | 12.54% |
| | Facil | 88.02% | 96.65% | 92.72% | 95.58% | 85.42% | 93.72% | 74.06% | 40.16% | 21.84% |
| | MobNet | 66.73% | 60.48% | 67.69% | 67.98% | 56.71% | 79.45% | 82.17% | 88.27% | 83.75% |
| ECONOMY | Earn | 978.10 | 918.10 | 1,052.50 | 934.80 | 1,184.00 | 968.20 | 999.00 | 1,065.40 | 1,096.40 |
| | FamInc | 15,061.00 | 15,774.00 | 18,743.00 | 15,546.00 | 16,512.00 | 15,958.00 | 16,025.00 | 17,484.00 | 17,337.00 |
| | PurcPw | 79.65% | 79.55% | 93.69% | 78.49% | 92.45% | 85.92% | 99.10% | 87.29% | 86.51% |
| | Unemp | 11.84% | 10.87% | 10.27% | 13.18% | 10.90% | 15.66% | 15.74% | 11.13% | 14.65% |
| | IncPrim | 104.73 | 111.06 | 317.81 | 116.53 | 520.11 | 258.57 | 312.10 | 346.32 | 85.53 |
| | PrimGVA | 2.20% | 10.81% | 2.72% | 4.01% | 20.95% | 8.64% | 3.89% | 9.24% | 1.52% |
| | R&D | 0.54% | 0.76% | 2.24% | 1.08% | 0.10% | 0.40% | 0.30% | 0.30% | 0.36% |
| ENVIRONMENT | RenEn | 0.21% | 0.15% | 0.20% | 0.16% | 0.21% | 0.11% | 0.38% | 0.25% | 1.43% |
| | WasteW | 100.00% | 77.56% | 90.33% | 100.00% | 100.00% | 90.00% | 100.00% | 46.00% | 87.00% |
| | ProtectA | 16.00% | 24.40% | 0.19% | 19.00% | 10.30% | 9.30% | 9.40% | 24.20% | 58.20% |
| | WatQlt | 75.20% | 68.00% | 45.00% | 40.70% | 54.90% | 26.90% | 76.20% | 65.00% | 65.00% |
| | SustAgr | 0.49% | 2.49% | 0.20% | 5.78% | 0.34% | 3.95% | 0.89% | 0.14% | 2.15% |
| | ExpHer | 65.53 | 111.17 | 67.34 | 84.97 | 98.19 | 142.36 | 138.59 | 50.16 | 118.67 |

SOURCE: Own elaboration, from Instituto Nacional de Estatística de Portugal and (Abreu and Mesias, 2020).

From the previous table, some relevant aspects can be observed that may help to understand the subsequent results. Thus, Trás-os-Montes is the most aged region and has the highest dependency index, while Madeira and Açores have the lowest values.

Within the social dimension, the region of Coimbra stands out as an industrialized area with strong urbanization cantered in the city of Coimbra, which is reflected in good indicators with respect to health services, labour force qualification or the percentage of people with university studies.

These aspects are reflected in the economic pillar. Thus, although the highest per capita incomes are found in Alentejo Litoral, Madeira and Açores (probably due to the high impact of tourism in the economy of these areas), Coimbra is the region with the highest income per family unit, the lowest unemployment rate (which is reflected in a higher number of people working in the family units and in the income per family) and the highest expenditure on R&D.

Finally, and regarding the environmental dimension, the urban and industrial character of Coimbra is also noteworthy, which is reflected in its data on protected areas, water quality or sustainable agriculture, which are among the worst of the regions analysed.

Subsequently, all the data were normalized, a procedure required prior to any data aggregation, as the indicators often have different measurement units (OECD, 2008) and it wouldn't be possible to comparable them otherwise. A range of [0-1] was used to normalize the variables, although an "inverse" normalization was applied to the indicators *Demographic Dependency Index, Unemployment rate* and *Proportion of Population Aged 65 or over* by assigning the value of 0 to the NUTS III with the highest values in these indicators and 1 to those with the lowest values, as applied by other authors (Kynčlová et al., 2020).

Table 3 presents the normalized results of the three RD Indexes obtained by applying the aggregation methods explained in the Methodology section to the data presented in table 2.

**Table 3. RDI results with the different 3 methodologies for the NUTS III regions analysed**

| | RDI Abreu | RDI Delphi | RDI PCA |
|---|---|---|---|
| Alto Minho | 0.34 | 0.13 | 0.29 |
| Terras de Trás-os-Montes | 0.00 | 0.11 | 0.00 |
| Região de Coimbra | 0.78 | 0.89 | 0.27 |
| Beiras e Serra da Estrela | 0.05 | 0.02 | 0.01 |
| Alentejo Litoral | 0.37 | 0.38 | 0.18 |
| Alto Alentejo | 0.17 | 0.00 | 0.09 |
| Algarve | 0.94 | 0.84 | 0.70 |
| Região Autónoma dos Açores | 0.74 | 0.73 | 1.00 |
| Região Autónoma da Madeira | 1.00 | 1.00 | 0.99 |

SOURCE: Own elaboration.

The results in Table 3 clearly show that there is a strong relationship between *RDI Abreu* and *RDI Delphi*, with the NUTS III ranked similarly, even though *RDI Delphi* seems to penalize the least developed regions which get lower ratings than in *RDI Abreu*, and therefore, are comparatively worst positioned. In fact, using the Pearson's correlation coefficient, the most common measure of association between two continuous variables (Tabatabai et al., 2021), the value obtained between the *RDI Abreu* and *RDI Delphi* indexes ($r$=0.96) suggests that these two methods generate very similar results, almost as when there is a perfect positive linear relation (Sari et al., 2017)[2]. On the other hand, the Pearson's values calculated between *RDI Abreu/RDI PCA* and *RDI Delphi/RDI PCA* are, respectively, $r$=0.86 and $r$=0.80, meaning that they are both feasible methodologies, as they have a high positive correlation with an accurate method as it is *PCA*.

As a complementary analysis, the linear relationship between each pair of indices can be seen in Figure 2:

[2] The closer this value to zero, the smaller is the degree of linear relation (Sari et al., 2017).

**Figure 2. Scatter Plot Matrix**

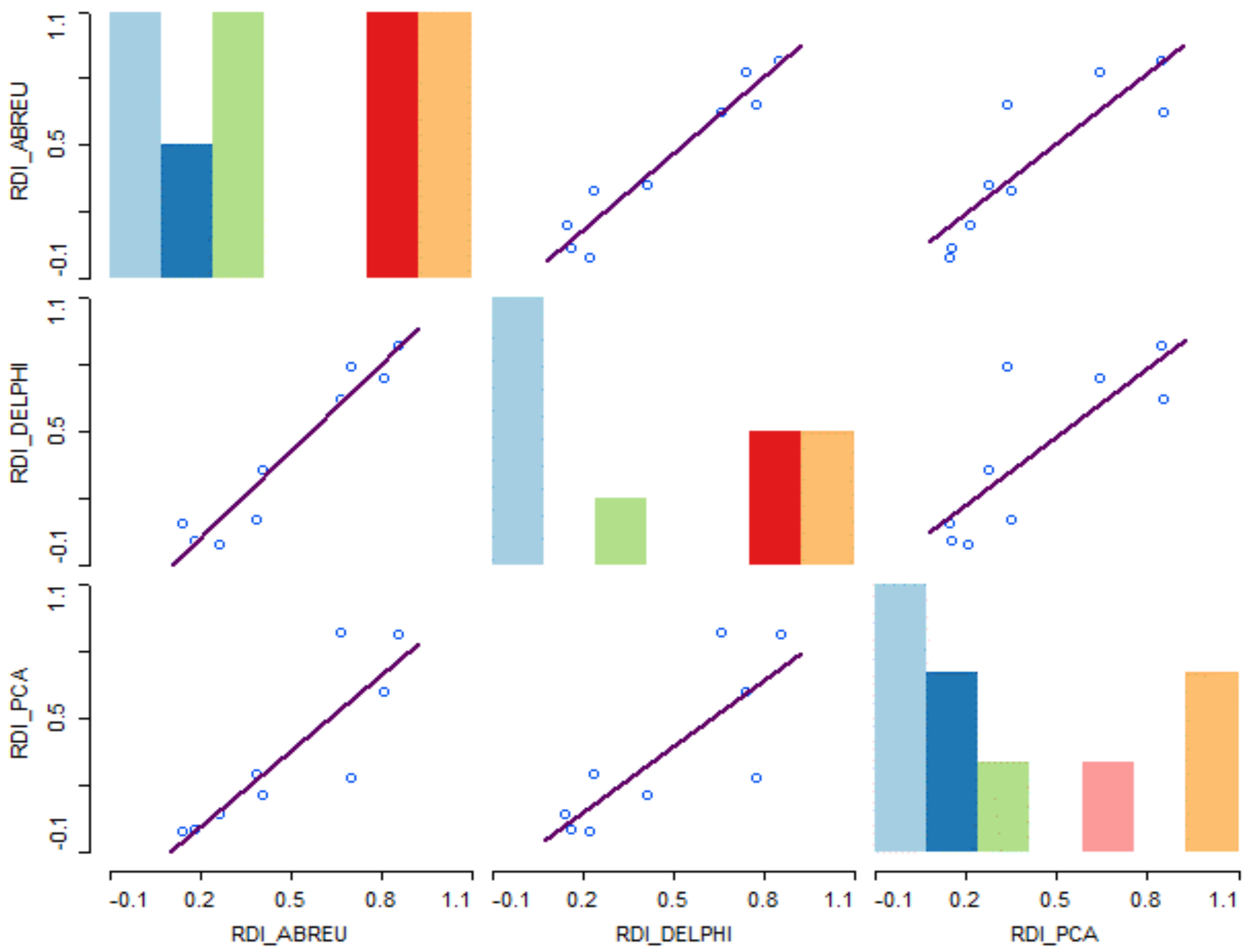

SOURCE: Own elaboration.

To display the distribution of the values between the three RDI indices and to obtain an indication of how these values are spread out, the results shown in table 3 were also represented in boxplots[3] in Figure 3, which displays the minimum (*min*), first quartile (*Q1*), mean and median, third quartile (*Q3*), interquartile range (*IQR*), standard deviation (*s.d.*), and the maximum (*max*) values:

[3] In a boxplot, a box is drawn from the first quartile (*Q1*) to the third quartile (*Q3*) and a vertical line goes through the box at the median (green circle). Also, the *Q1*-1,5**IQR* and *Q3*+1,5**IQR* values are represented as black horizontal lines, and the mean value is represented as a red line.

**Figure 3. Boxplots representing the three RDI values regarding the 3 methods**

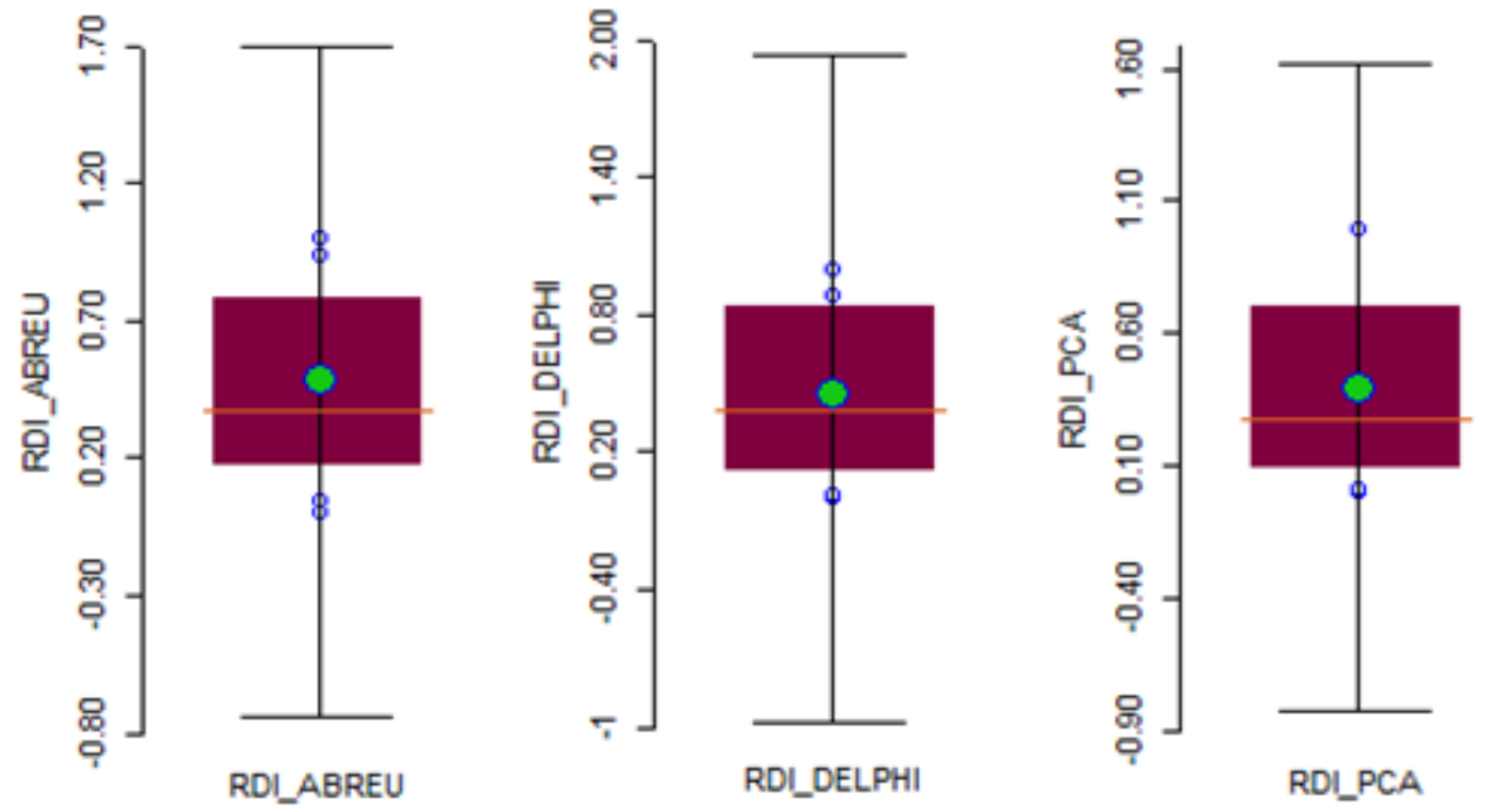

| | RDI_ABREU | RDI_DELPHI | RDI_PCA |
|---|---|---|---|
| min | 0 | 0 | 0 |
| max | 1 | 1 | 1 |
| Q1 | 0.1700 | 0.1100 | 0.0900 |
| median | 0.3700 | 0.3800 | 0.2700 |
| Q3 | 0.7800 | 0.8400 | 0.7000 |
| IQR | 0.6100 | 0.7300 | 0.6100 |
| mean | 0.4878 | 0.4556 | 0.3922 |
| s.d. | 0.3844 | 0.4088 | 0.4005 |

SOURCE: Own elaboration.

From the information presented in Figure 3 we can conclude that there are no major differences in the range and distribution of the three calculated indices. The index with the greatest variability is *RDI Delphi* (*IQR*=0.73, *s.d.*=0.41), and it is also observed that the highest average value corresponds to *RDI Abreu* (*mean*=0.49) while the lowest median value is that of *RDI PCA* index (*median*=0.27).

However, the NUTS III with the lowest value is different when applying *RDI Abreu* and *RDI PCA* (Terras de Trás-os-Montes), or *RDI Delphi* (Alto Alentejo), although in fact both regions are strongly needed for public policies to reverse depopulation and aging, which leverage their development.

Regarding the NUTS III with the highest value, both *RDI Abreu* and *RDI Delphi* grant the highest rating to the Região Autónoma da Madeira. However, when applying *RDI PCA* this is the territory with the second highest value, giving the first place to Região Autónoma dos Açores, but with a result really close to the first one.

Nevertheless, although the results are similar in the extreme values, the situation is different for the middle ones. For example, when analysing Região de Coimbra, *RDI Abreu* and *RDI Delphi* rank it in the third and second place, respectively, but the *RDI PCA* positions this NUTS III in the fifth position, after Alto Minho. This result raises some doubts about its reliability, as Coimbra is one of the most industrialized and developed regions of Portugal. In fact, *RDI Abreu* and *RDI Delphi* rank Alto Minho on the seventh place, a figure that matches with a low-density territory, strongly marked by aging.

These conclusions are again verified through the use of a parallel coordinates plot (Figure 4), one of the most popular techniques for visualization and analysis of multidimensional data (Fu et al., 2016). On Figure 4 we can observe the different RDI results presented by the three methodologies with the relationships that exists between them: comparing the position of the 9 NUTS III RDI values generated by *RDI Abreu* and *RDI Delphi*, the lines mostly don't cross each other, meaning that there is a strong relation between the different values. But when we introduce the results of *RDI PCA*, we can see that the situation is very different, with a lot of changes in its relative positions.

**Figure 4. Parallel coordinates plot**

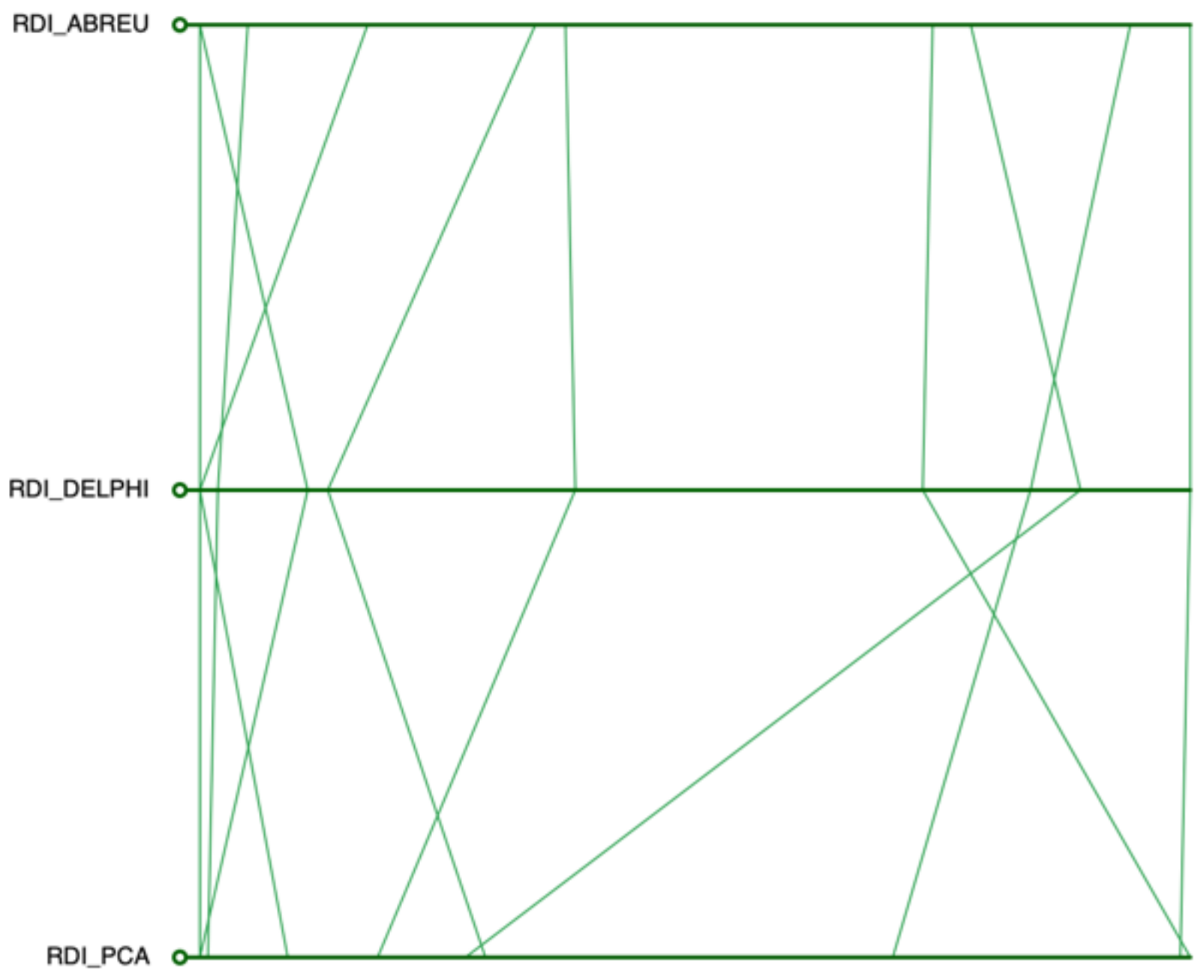

SOURCE: Own elaboration.

### 3.2. Assessment by experts

In the final part of this study, the group of experts received the information presented in Table 3 together with a ranking of the regions prepared from the same table. They were also provided with an introductory text where the issue of developing indexes to measure rural development was briefly presented, together with an explanation of the selection of the 9 NUTS III regions. After this information, the experts were asked to assess each of the RDI methods:

"*On a scale of 1 to 5 and taking into account the results presented to you, how adequate do you consider each of the methods to be for measuring rural development, with 1 being "not at all adequate" and 5 being "very adequate"?*"

They were also asked to indicate which of the results had been most relevant for their answers. Table 4 presents the ratings granted for the experts to each of the three aggregation methods.

**Table 4. Experts' adequacy rating for the different 3 aggregating methodologies (1: not at all adequate; 5: very adequate)**

| Expert | RDI Abreu | RDI Delphi | RDI PCA |
|---|---|---|---|
| 1, 13, 16 | 3 | 3 | 3 |
| 2, 10, 25 | 4 | 4 | 2 |
| 3 | 4 | 2 | 3 |
| 4 | 4 | 1 | 3 |
| 5 | 3 | 3 | 4 |
| 6 | 2 | 4 | 1 |
| 7 | 5 | 5 | 5 |
| 8 | 4 | 3 | 4 |
| 9 | 3 | 4 | 4 |
| 11 | 1 | 4 | 3 |
| 12 | 4 | 4 | 3 |
| 14 | 4 | 4 | 4 |
| 15 | 5 | 3 | 3 |
| 17 | 5 | 1 | 1 |
| 18 | 3 | 2 | 2 |
| 19, 21 | 3 | 4 | 2 |
| 20, 22 | 4 | 3 | 2 |
| 23 | 3 | 5 | 4 |
| 24 | 3 | 4 | 3 |

SOURCE: Own elaboration.

The results in Table 4 show that the most adequate aggregating methodology from the point of view of the experts is *RDI Abreu*, which has got a total score of 88 points (median=4) followed by *RDI Delphi* with a score of 84 points (median=4). The worst-scoring methodology is the one based on the weighting granted to the different indicators through the PCA (*RDI PCA*), which scored 70 points (median=3).

Considering the ranking, the difference between *RDI Abreu* and *RDI Delphi* is minimal, (17 experts granted the first position to *RDI Abreu* vs 16 to *RDI Delphi*; 6 experts granted the 2$^{nd}$ position to *RDI Abreu* vs 7 to *RDI Delphi*).

**Table 5. Ranking granted by the experts to the three RD indexes**

| Ranking position | RDI_ABREU | RDI_DELPHI | RDI_PCA |
|---|---|---|---|
| 1º | 17 | 16 | 8 |
| 2º | 6 | 7 | 12 |
| 3º | 2 | 2 | 5 |

SOURCE: Own elaboration.

When analyzing the reasons provided by the experts to support their decisions, the main points of disagreement with *RDI PCA* come from the relative position of the regions of Coimbra and Açores. The former gets the highest rating in *RDI PCA*, while Coimbra goes down to the fifth position below Alto Minho region, raising some doubts between the experts about its reliability. In fact, Coimbra, according to 2018 data from INE, is one of the most industrialized and developed regions of Portugal, whereas Alto Minho is a low-density territory, strongly marked by aging (Conselho Estratégico de Desenvolvimento Intermunicipal, 2021). On the other hand, Açores is the Portuguese region with one of the worst health services indicator and highest income inequality (Diogo, 2019), therefore its top position on *RDI PCA* is not well accepted by the experts.

The result of the experts' assessment is consistent with the more usual trend in composite index construction (Greco et al., 2019; OECD, 2008). Furthermore, it provides a basis for the use of those indexes which have considered that the different dimensions of RD have similar importance, such as those developed by (Ristić et al., 2019) or (Abreu et al., 2019). However, other indexes that rely on the use of statistical techniques to aggregate the variables composing the RD index, or to determine their weight (Bolcárová and Kološta, 2015; Kiryluk-Dryjska and Beba, 2018; Ma et al., 2020), but where no validation of their accuracy has been performed, should rethink their formulation in view of the results presented in this paper.

## 4. CONCLUSIONS

Despite the relevance of rural development for current policies and society, the lack of indexes specifically designed for its evaluation in rural areas hinders the design and implementation of policies, or the adjustment of those that, being already implemented, may not generate the expected results. It also makes it difficult to identify areas where the use of public funds would be more effective, resulting in less efficient rural development policies.

In addition, the various approaches used in the design of those indexes tend to suffer from a lack of stakeholder involvement, both in the selection of the indicators to be included in the study and in the different ways of combining them to generate a unit of measurement.

The approach taken in this paper aims to bridge these gaps by using a series of indicators previously generated by a panel of rural development experts, which have become the inputs to generate several rural development indices by applying different aggregation methodologies (arithmetic/geometric mean with and without weighting -*RDI Abreu* and *RDI Delphi*) and aggregation by means of principal component analysis -*RDI PCA*).

The results have shown the statistical similarity of two of the proposed indexes (*RDI Abreu* and *RDI Delphi*) compared to the one obtained by *RDI PCA*. Finally, a panel of rural development experts have reviewed the scores generated by the three indexes for a set of 9 NUTS III Portuguese regions. *RDI Abreu* was considered to reflect more accurately the different levels of rural development, while the experts disagreed with some of the results generated by *RDI PCA*. Thus, this work offers an objective rationale for the definition of indexes used to assess rural development, which can be widely used in a context where the rural environment and the policies that promote its development are gaining more and more importance.

## 5. CREDIT AUTHORSHIP CONTRIBUTION STATEMENT

Isabel Abreu: Conceptualization, Visualization, Validation, Methodology, Investigation, Formal analysis, Writing - original draft. Francisco J. Mesías: Conceptualization, Visualization, Validation, Methodology, Supervision, Writing - review & editing. Julián Ramajo: Methodology, Formal analysis, Software, Supervision, Writing - review & editing.

## 6. DECLARATION OF INTEREST

None

## 7. REFERENCES

Abreu, I., 2014. Construção de um índice de desenvolvimento rural e sua aplicação ao Alto Alentejo. Instituto Politécnico de Portalegre.

Abreu, I., Mesias, F.J., 2020. The assessment of rural development: Identification of an applicable set of indicators through a Delphi approach. J. Rural Stud. https://doi.org/10.1016/j.jrurstud.2020.10.045

Abreu, I., Nunes, J.M., Mesias, F.J., 2019. Can Rural Development Be Measured? Design and Application of a Synthetic Index to Portuguese Municipalities. Soc. Indic. Res. 145, 1107–1123. https://doi.org/10.1007/s11205-019-02124-w

Annoni, P., Dijkstra, L., 2019. The EU Regional Competitiveness Index 2019. Eur. Comm. 1–42.

Bélanger, V., Vanasse, A., Parent, D., Allard, G., Pellerin, D., 2012. Development of agri-environmental indicators to assess dairy farm sustainability in Quebec, Eastern Canada. Ecol. Indic. 23, 421–430. https://doi.org/10.1016/j.ecolind.2012.04.027

Benitez-Capistros, F., Hugué, J., Koedam, N., 2014. Environmental impacts on the Galapagos Islands: Identification of interactions, perceptions and steps ahead. Ecol. Indic. 38, 113–123.

https://doi.org/10.1016/j.ecolind.2013.10.019

Bolcárová, P., Kološta, S., 2015. Assessment of sustainable development in the EU 27 using aggregated SD index. Ecol. Indic. 48, 699–705. https://doi.org/10.1016/j.ecolind.2014.09.001

Carraro, C., Cruciani, C., Ciampalini, F., Giove, S., Lanzi, E., 2009. Aggregation and projection of sustainability indicators: a new approach, in: 3rd OECD World Forum on "Statistics, Knowledge and Policy" Charting Progress, Building Visions, Improving Life.

Conceição, P., 2019. Human Development Report 2019: beyond income, beyond averages, beyond today, United Nations Development Program.

Conselho Estratégico de Desenvolvimento Intermunicipal, 2021. Alto Minho 2030. A Construção da Estratégia Alto Minho.

Davis, G., 2004. A History of the Social Development Network in The World Bank, 1973 - 2002, Social Development. Washington, D.C.

Decancq, K., Lugo, M.A., 2008. Setting Weights in Multidimensional Indices of Well-being and Deprivation, OPHI Working Paper 18. https://doi.org/10.2307/1213275

DEFRA, 2004. Regional quality of life counts-2003. Regional versions of the headline indicators of sustainable development. Department for Environment, Food and Rural Affairs, London.

Diogo, F., 2019. Algumas Peculiaridades Da Pobreza Nos Açores. Sociol. Line 2018, 81–101. https://doi.org/10.30553/sociologiaonline.2019.19.4

Escribano, M., Díaz-Caro, C., Mesias, F.J., 2018. A participative approach to develop sustainability indicators for dehesa agroforestry farms. Sci. Total Environ. 640–641. https://doi.org/10.1016/j.scitotenv.2018.05.297

European Commission, 2017. The Future of Food and Farming. Brussels.

Eurostat, 2019. Eurostat Regional Yearbook. 2019 edition. Publications Office of the European Union, Luxembourg. https://doi.org/10.27585/1522

Fu, L., Lin, M., Zhang, J., 2016. Journal of Visual Languages and Computing Two axes re-ordering methods in parallel coordinates plots. J. Vis. Lang. Comput. 33, 3–12. https://doi.org/10.1016/j.jvlc.2015.12.001

Governo Regional dos Açores, 2021. Estado das massas de água da Região Hidrográfica dos Açores (RH9) [WWW Document]. Relatório do Estado do Ambient. dos Açores. URL http://rea.azores.gov.pt/reaa/54/agua/875/estado-das-massas-de-agua-da-regiao-hidrograf (accessed 3.1.21).

Greco, S., Ishizaka, A., Tasiou, M., Torrisi, G., 2019. On the Methodological Framework of Composite Indices: A Review of the Issues of Weighting, Aggregation, and Robustness. Soc. Indic. Res. 141, 61–94. https://doi.org/10.1007/s11205-017-1832-9

Haag, A., 2009. Performance of the National Program for Strengthening Family Agriculture in the State of Rio Grande do Sul. Universidade Federal do Rio Grande do Sul.

Horrillo, A., Escribano, M., Mesias, F.J., Elghannam, A., Gaspar, P., 2016. Is there a future for organic production in high ecological value ecosystems? Agric. Syst. 143, 114–125. https://doi.org/10.1016/j.agsy.2015.12.015

Jollife, I.T., Cadima, J., 2016. Principal component analysis: A review and recent developments. Philos. Trans. R. Soc. A Math. Phys. Eng. Sci. https://doi.org/10.1098/rsta.2015.0202

Kageyama, A., 2008. Desenvolvimento rural : conceitos e aplicação ao caso brasileiro. UFRGS Editora, Porto Alegre (Brasil).

Kiryluk-Dryjska, E., Beba, P., 2018. Region-specific budgeting of rural development funds—An application study. Land use policy 77, 126–134.

https://doi.org/10.1016/j.landusepol.2018.05.029

Kynčlová, P., Upadhyaya, S., Nice, T., 2020. Composite index as a measure on achieving Sustainable Development Goal 9 (SDG-9) industry-related targets: The SDG-9 index. Appl. Energy 265. https://doi.org/10.1016/j.apenergy.2020.114755

Ma, L., Liu, S., Fang, F., Che, X., Chen, M., 2020. Evaluation of urban-rural difference and integration based on quality of life. Sustain. Cities Soc. 54, 101877. https://doi.org/10.1016/j.scs.2019.101877

Michalek, J., Zarnekow, N., 2012a. Construction and application of the Rural Development Index to analysis of rural regions. Luxembourg.

Michalek, J., Zarnekow, N., 2012b. Application of the Rural Development Index to Analysis of Rural Regions in Poland and Slovakia. Soc. Indic. Res. 105, 1–37. https://doi.org/10.1007/s11205-010-9765-6

OECD, 2008. Handbook on Constructing Composite Indicators, OECD Publications.

Oxford Poverty and Human Development Initiative, 2010. Multidimensional Poverty Index.

Ristić, D., Vukoičić, D., Milinčić, M., 2019. Tourism and sustainable development of rural settlements in protected areas - Example NP Kopaonik (Serbia). Land use policy 89, 104231. https://doi.org/10.1016/j.landusepol.2019.104231

Sari, B.G., Lúcio, A.D., Santana, C.S., Krysczun, D.K., Tischler, A.L., Drebes, L., 2017. Sample size for estimation of the Pearson correlation coefficient in cherry tomato tests. Ciência Rural 47, 1–7. https://doi.org/10.1590/0103-8478cr20170116

Tabatabai, M., Bailey, S., Bursac, Z., Tabatabai, H., Wilus, D., Singh, K.P., 2021. An introduction to new robust linear and monotonic correlation coefficients. BMC Bioinformatics 22, 1–18. https://doi.org/10.1186/s12859-021-04098-4

UNDP, 2016. Human development report 2016, United Nations Development Programme. https://doi.org/eISBN: 978-92-1-060036-1

UNDP, 2010. Human Development Report 2010 The Real Wealth of Nations : Pathways to Human Development, Human Development. https://doi.org/10.2307/2137795

UPA, 2016. Desarrollo rural. Oportunidades desaprovechadas. La Tierra 254, 31–33.

World Bank, 1997. Expanding the measure of wealth. Indicators of environmentally sustainable development. World Bank, Washington, D.C.

Yilmaz, B., Daşdemir, I., Atmiş, E., Lise, W., 2010. Factors affecting rural development in turkey: Bartin case study. For. Policy Econ. 12, 239–249. https://doi.org/10.1016/j.forpol.2010.02.003